\documentclass[aps,prb,twocolumn,superscriptaddress,showpacs,floatfix,amsmath,amssymb]{revtex4}
\usepackage{graphicx}
\usepackage{color}
\usepackage{xcolor,soul}
\usepackage[normalem]{ulem}

\DeclareMathAlphabet\mathbfcal{OMS}{cmsy}{b}{n}

\newcommand{\br}[0]{ {\bf r} }

\newcommand{\FP}[0]{ \hat{\Psi} }

\newcommand{\KS}[0]{ {\rm {\tiny \tiny KS}} }

\newcommand{\bJ}[0]{ {\bf J} }
\newcommand{\bA}[0]{ {\bf A} }
\newcommand{\bj}[0]{ {\bf j} }
\newcommand{\vbJ}[0]{ {\vec {\bf J}} }
\newcommand{\vbA}[0]{ {\vec {\bf A}} }
\newcommand{\vA}[0]{ {\vec {A}} }
\newcommand{\bG}[0]{ {\bf G} }
\newcommand{\vbG}[0]{ {\vec {\bf G} } }

\usepackage{color}
\usepackage{booktabs}
\usepackage{longtable}
\usepackage{bm}

\renewcommand{\v}[1]	{\ensuremath{\bm{#1}}} 
\newcommand{\mr}[1]     {\ensuremath{\mathrm{#1}}}

\newcommand{\xc}[0]{ {\rm xc}  }
\newcommand{\EH}[0]{ E_{  {\rm H}}}

\newcommand{\Exc}[0]{ E_\xc }

\begin{document}

\title{Spin-currents via the gauge-principle for meta-generalized-gradient exchange-correlation functionals}

\author{Jacques K. Desmarais}
\email{jacqueskontak.desmarais@unito.it}
\affiliation{Dipartimento di Chimica, Universit\`{a} di Torino, via Giuria 5, 10125 Torino, Italy}

\author{Jefferson Maul}
\affiliation{Dipartimento di Chimica, Universit\`{a} di Torino, via Giuria 5, 10125 Torino, Italy}

\author{Bartolomeo Civalleri}
\affiliation{Dipartimento di Chimica, Universit\`{a} di Torino, via Giuria 5, 10125 Torino, Italy}

\author{Alessandro Erba}
\affiliation{Dipartimento di Chimica, Universit\`{a} di Torino, via Giuria 5, 10125 Torino, Italy}

\author{Giovanni Vignale}
\affiliation{Institute for Functional Intelligent Materials, National University of Singapore, 4 Science Drive 2, Singapore 117544}

\author{Stefano Pittalis}
\email{stefano.pittalis@nano.cnr.it}
\affiliation{Istituto Nanoscienze, Consiglio Nazionale delle Ricerche, Via Campi 213A, I-41125 Modena, Italy}

\date{\today}

\begin{abstract}
The prominence of density functional theory (DFT) in the field of electronic structure computation stems from its ability to  usefully balance accuracy and computational effort. 
At the base of this ability  is a functional of the electron density: the exchange-correlation  energy.
This functional satisfies known exact conditions that  guide the derivation of approximations. 
The strongly-constrained-appropriately-normed (SCAN)  approximation stands out as a successful, modern,  example.
In this work, we demonstrate how the SU(2) gauge-invariance of the 
exchange-correlation functional in spin current density functional theory allows us 
to add an explicit dependence on spin currents in the SCAN functional (here called JSCAN) --- and similar meta-generalized-gradient functional approximations --- solely invoking first principles. 
In passing, a spin-current dependent generalization of the electron localization function (here called JELF) is also derived. The extended forms are implemented in a developer's version of the \textsc{Crystal23} program. Applications on molecules and materials confirm  the practical  relevance of the extensions.
\end{abstract}

\pacs{71.15.Mb, 71.15Rf, 31.15.E-}

\maketitle

{\em Introduction}.
Spin current carrying states have acquired a major role in condensed matter physics and electronic device engineering. \cite{greiner2012relativistic} 
Recently, the interest in spin currents has  grown further following the discovery of a host of materials which exhibit nontrivial topological properties \cite{bernevig2013topological,vanderbilt2018berry}  and  the development of new techniques to accurately control the spin of the electrons in spintronic and quantum computing devices.\cite{RevModPhys.76.323}

Commonly, spin currents originate from spin-orbit coupling (SOC). In many situations, SOC can be accounted for via the two-component Pauli equation, which offers formal and computational simplification as compared to the full Dirac equation. Just as an electric potential couples to the particle density and a magnetic field couples to the spin  and to the particle (i.e. ordinary) current, SOC can be described as a tensor field which couples to the spin current (SC), i.e., the tensor product of spin and ordinary current.   Unlike the ordinary current, the spin current is invariant under time reversal (being the product of two time-reversal odd quantities) and therefore can freely flow in the time-reversal invariant ground state of a non magnetic system.\cite{Tokatly2008,Droghetti2022}

It is sometimes adequate (for systems composed of light atoms)  to include SOC as a small perturbation  on top of the Kohn-Sham (KS) potential of ordinary density functional theory (DFT).\cite{Kohn1999} This approach can in principle   find the   particle density and total energy of  real {\em interacting}  ground states by solving a single-particle problem  in the presence of a self-consistent exchange-correlation (xc) potential, which is a functional of the density. Also, the eigenvalues of the KS single-particle equation provide a useful representation of the band structure and related properties.\cite{Seidl1996,perdew2017understanding,NazarovVignale20111,Prokopiou2022} 

However, there is no reason why a pure density functional approximation that was developed for systems lacking spin currents would still yield accurate results when   SOC is present.  To accomplish this more ambitious task --- yet within the two-component Pauli equation --- we need an exchange-correlation energy functional that depends not only on the density but also --- and crucially --- on the spin current density, and does so in a universal (i.e., system-independent) manner, as mandated by the original spin-current density functional theory (SCDFT)\cite{VignaleRasolt:88, Bencheikh:03} and its more recent developments. \cite{Goerling2006,HeatonBurgess2007,Goerling2018,desmarais2024generalized}  In practice, this means including in the effective potential, not only the ``bare" SOC, but also its ``dressing" due to many-body effects, in analogy to the dressing of the bare electron-nuclear potential by Hartree and exchange-correlation (xc) effects.

Crucially, any xc energy functional that depends on the spin current density must be invariant --- not only under U(1) gauge transformations\cite{VignaleRasolt:87,TP05} --- but also under SU(2) transformations\cite{VignaleRasolt:88, Bencheikh:03,Pittalis2017}. The latter can be viewed as local rotations of the spinorial state of the electrons\cite{FroehlichStuder:93,Jin_2006}. Failing to satisfy SU(2) invariance would have disastrous consequences for a density functional approximation: it would mean that the approximation does not  distinguish between actual variations of the spin current density --- which can have physical consequences --- and variations arising from gauge transformations, which must have no consequence! Thus the gauge invariance of the exact xc-energy functional acts as {a} most important exact constraint, which guides the construction of approximate functionals.

While different approximate functionals obviously have different strengths and weaknesses, \emph{they all must satisfy the constraint of gauge-invariance.} Hence, we focus on the strongly-constrained-appropriately-normed (SCAN) functional,\cite{SCAN} which is currently recognized as the least  empirical general-purpose approximation in the class of Meta-Generalized-Gradient Approximations (MGGA).\cite{Workhorse2009,CMW2013,SCAN,CHEMSCAN,Fabien2016,Peng2016,Brandenburg2016,Isaacs2018,Tran2019} However, our procedure is applicable to any approximate functional in this class.

The exceptional performance of the SCAN is fundamentally due to its ability to satisfy altogether a large number of known exact conditions for the xc-energy functional of DFT.
Furthermore, among its crucial ingredients, the KS kinetic-energy density appears in a combination of variables that was originally proposed for the so-called electron localization function (ELF) -- a quantity that 
provides a vivid visualization of molecular bonds and atomic shells (more below).\cite{BE90,Savin94}

But regular DFT  cannot deal with spin-currents induced by extra interactions.
This brings forth the necessity to switch to SCDFT both in principle\cite{VignaleRasolt:88, Bencheikh:03} and in practice.\cite{Goerling2006,HeatonBurgess2007,Goerling2018,desmarais2024generalized,desmarais2020adiabatic,bodo2022spin,comaskey2022spin} 
The dependence of the xc energy functional on spin currents in SCDFT is
strongly constrained by the requirement that its form remains invariant under local SU(2)
gauge transformations (defined below). Taking as a starting point the SCAN and the ELF forms, in this work  we show how to make them functionals of the spin currents while preserving the fundamental SU(2) gauge invariance of the theory. The resulting generalizations are here called JELF and JSCAN. A set of computed results for molecules and materials demonstrates the practical usefulness of our non-empirical extension of the SCAN functional. An implementation is presented in a developer's version of the \textsc{Crystal23} program.\cite{erba23cry}

{\em SCDFT in a nutshell}. 
We shall consider ground states at geometries that do not magnetize  nor host particle currents spontaneously. 
Furthermore, magnetic couplings (Zeeman field, Abelian vector potential) that  break time-reversal symmetry are not included.
Accounting for  non-Abelian vector potentials, however, we must consider together with the particle density, $n$, the spin currents, $\vbJ$.\cite{FroehlichStuder:93,Bencheikh:03,AbedinpourTokatly:10,ZG16,Droghetti2022}
The energy density functional of interest, thus, reads as follows:
\begin{widetext}
 \begin{eqnarray}\label{H_1}
E_{v, \vbA}[n,\vbJ ] = T_s[n, \vbJ] + \EH[n] + \Exc[n,\vbJ]  +    \frac{1}{c} \int d^3r~    \bJ^a \cdot \bA^a 
+ \int d^3r~ {n} \left(v + \frac{1}{2c^2 } \bA^a  \cdot \bA^a \right)
\end{eqnarray}
\end{widetext}
where $T_s$ is the kinetic energy of the KS system, while $\EH$ and $\Exc$ are Hartree and xc energies.
Here and in the following, we denote with bold characters, $\boldsymbol{v}$, quantities with spatial indices (Greek lower indices, $v_\mu$, when written explicitly); and use an arrow, $\vec{v}$, to denote quantities with spin indices (upper Latin indices, $v^a$, when written explicitly)  --- $\mu$ and $a$ have values $x$,$y$,$z$. Therefore $\vbJ$ and $\vbA$ have both spatial and spin indices. Contractions over spatial indices are denoted with ``$\cdot$'' and Einstein convention is used for summing over repeated indices.
Unless otherwise stated, we use Hartree atomic units in which $\hbar = m = 1$.
The corresponding KS equations are:
\begin{widetext}
\begin{eqnarray}
\label{GKS-RES}
\left\{ \frac{1}{2}\left[ -i  \nabla + \frac{1}{c} {\mathbfcal{A}}_\KS \right]^2  +  \left[  
 \left( v  + v_{\rm Hxc} \right)
 + \frac{1}{2c^2} \left( \mathbfcal{A}^2-  {\mathbfcal A}^2_\KS  \right) \right] \right\} \Phi_\kappa = \varepsilon_\kappa \Phi_\kappa \;
\end{eqnarray}
\end{widetext}
where 
$
{ \mathbfcal{A}_\KS } =  { \mathbfcal{A} } + { \mathbfcal{A}_\xc }  = {\sigma}^a  {\bA}^a  + {\sigma}^a {\bA}_{\rm xc}^a
$, with $\vec{\sigma}$ being the vector of Pauli matrices,
$ v_{\rm Hxc} = \delta E_{\rm Hxc}/\delta n $ is a Hartree-xc-scalar potential (analogous to the Hxc-scalar potential of DFT)
and
$
{\frac{1}{c}}{\vbA}_{\rm xc} = \delta E_{\rm xc} / \delta \vbJ
$
is a non-Abelian xc-vector potential (which is specific to SCDFT).
The occupied (lowest in energy)  two-component KS spinors $\Phi_k$ allow us to compute the particle density,
$n  = \sum_{k=1}^{N} \Phi^\dagger_k \Phi_k^{\vphantom{\dagger}}$, and
the spin current
$
\vbJ = \frac{1}{2i} \sum_{k=1}^{N} \Phi^\dagger_k \vec{\sigma} \left[  \nabla \Phi_k^{\vphantom{\dagger}}\right] -  \left[ \nabla \Phi^\dagger_k \right]  \vec{\sigma} \Phi_k^{\vphantom{\dagger}}$
of the real state and, thus, the energy as well. All this is exact in principle, while, in practice, we are bound to approximate $\Exc$.

Invoking a MGGA, it is  convenient to switch to the so-called generalized-KS (GKS) approach to
SCDFT.\cite{desmarais2024generalized}
This means that for  generating the single-particle equations,
we differentiate the energy functional expression explicitly w.r.t. the orbitals rather than 
w.r.t. to the densities~\cite{note2} 
--- nowadays, this has become a standard procedure in DFT.\cite{neumann1996}

A central property of the xc-energy functional in SCDFT is its form invariance under local SU(2) transformations, i.e., transformations that
act on the KS spinors as follows:
$\Phi(\br) \rightarrow  U_S(\br)  \FP(\br) $, where $U_S(\br) = \exp\big[\frac{i}{c} {\sigma}^a {\lambda}^a (\br)\big]$,
i.e., subjecting them to different rotations at different points in space.  
This invariance is expressed by the equation
\begin{align}\label{eq:FI}
E_{\rm xc}[n', {\vbJ'}] = E_{\rm xc}[n,\vbJ]
\end{align}
where 
$
n  \to n'=n$, $\bJ^a \to {\bJ'}^{a} = R^{ab} \big[\bJ^b + \frac{n}{c} \bG^b  \big]
$, 
$R^{ab}$ is a $3 \times 3$ matrix describing a rotation in $\mathbb{R}^3$ around the direction $\hat{\lambda}$ by an angle 
$\varphi=-2\lambda / c$, and
$\vbG = -\frac{i c}{2} {\rm Tr} \left( \vec{\sigma} U^{\dagger}_{\rm S} \nabla  U_{\rm S}^{\vphantom{\dagger}}  \right)$. \cite{note3} 

It is important to notice that  $\vbJ$ is not a gauge invariant quantity since it can appear 
as a consequence of a generic SU(2) transformation acting on a state in which $\vbJ=0$.  The invariance of  $E_{\rm xc}$ stated in Eq. \eqref{eq:FI}, implies that only certain gauge-invariant combinations of $\vbJ$  
with other KS quantities can legitimately appear in $E_{xc}$.  We now tackle the task of generalizing the SCAN functional (and the closely related ELF) to include a dependence on spin currents that complies  with Eq.~\eqref{eq:FI}.

{\em Extension of SCAN and ELF to spin-current carrying states}.
Recent works have moved valuable steps in developing extended DFT approaches to non-collinear magnetism and SOC.\cite{Pittalis2017,
Holzer2022,tancogne2023constructing,desmarais2024generalized} But  Ref.~\onlinecite{Pittalis2017} does not carry out applications, Ref.~\onlinecite{Holzer2022} does not work with full SU(2)-invariant xc-functionals, Ref.~\onlinecite{tancogne2023constructing} does not include an xc-vector potential in the solution, and Ref.~\onlinecite{desmarais2024generalized} does not include spin-current explicitly beyond exchange.

Here, we focus on MGGAs. The SCAN functional, for example, sets\cite{SCAN}
$E_\mr{xc}= \int d^3 r~ n(\br)  \epsilon^{\mr{SCAN}}_\mr{xc}(\br)$ where
$\epsilon_\mr{xc}^{\mr{SCAN}} = \epsilon_\mr{xc}^1 + \left(\epsilon_\mr{xc}^0 - \epsilon_\mr{xc}^1\right)f_\mr{xc}(\alpha)$
interpolates between the semi-local energy densities for single-orbital densities ($\epsilon_\mr{xc}^0$) and  for slowly-varying densities ($\epsilon_\mr{xc}^1$), which only depend on $n$ (a gauge-invariant quantity). The interpolation function $f_\mr{xc}$ is controlled by the  variable 
$\alpha = (\tau - \tau_{\rm W})/\tau_{\rm unif}$. 
The latter quantity entails three kinetic energy densities: the positive-definite conventional $\tau = 1/2\sum_i^{\mr{occ.}}|\nabla \phi_i|^2$ defined in terms of the occupied (one-component) KS orbitals \{$\phi_i$\}, the von Weizs\"acker $\tau_W = |\nabla n|^2/(8n)$ --- that is the {\em bosonic} expression of $\tau$ --- where $n=\sum_i^{\mr{occ.}}|\phi_i|^2$, and the kinetic energy of the noninteracting Fermi gas at uniform density $n$: $\tau_{\rm unif} = \frac{3}{10} (3\pi^2)^{2/3}\, n^{5/3}$.

The success of the SCAN is explained not only by the fact that it satisfies a large set of  known exact  conditions but also by the fact that it uses, specifically, $\alpha$ as a key variable. The $\alpha$ tends to zero in the iso-orbital limit (i.e., when the density is dominated by a single occupied orbital), tends to one in the uniform-density limit, and tends to infinity in regions dominated by
density overlap between closed shells.
In fact, $\alpha$ is a useful building block of density functional approximations,\cite{Becke1996,Workhorse2009,CMW2013} and it is the main ingredient of the
electron localization function,\cite{ELF90,Savin94,ELF97} ${\rm ELF} = 1 / (1 + \alpha^2)$ -- a simple but effective descriptor of molecular bonds and atomic shells.

More in general, the kinetic energy density must be expressed in terms of the occupied two-component KS spinors  
$\tau = 1/2\sum_k^{\mr{occ.}}  \nabla \Phi^{{\dagger}}_k \cdot \nabla \Phi_k^{\vphantom{\dagger}}$. 
Crucially, let us next consider the transformation of the kinetic energy under a local SU(2) transformation, $U_{\rm S}$, of the spinors.  
For the states considered in this work, one readily finds:~\cite{Pittalis2017}
$\tau  \to \tau' =\tau + \frac{1}{c} \bJ^a \cdot \bG^a + \frac{n}{2 c^2} \bG^a \cdot \bG^a$; for the definition of
$\vbG$ see below Eq. \eqref{eq:FI}.
As a consequence of which, it is apparent that $\alpha$ will change, too. 
{\em Hence, neither  the SCAN functional nor the ELF are SU(2)-form invariant.}
Such an invariance, however,   is not only an exact mathematical property of the xc-energy functional in SCDFT but --- because  
a local SU(2)  transformation may be regarded as part of an overall gauge transformation (i.e., a change in the description which cannot change the physics) --- it is also
a proper feature for any quantity that should carry direct chemical-physical information.
Fortunately, in one stroke, the substitution
$\tau  \to \widetilde{\tau} =\tau -   \frac{\bJ^a \cdot \bJ^a }{2n}$,
solves both issues: i.e., it introduces an explicit dependence on the spin currents and enforces the sought form invariance. In detail,
\begin{equation}\label{eq:talpha}
    \alpha \rightarrow  \widetilde{\alpha} = \alpha  - \frac{\bJ^a \cdot \bJ^a }{2n \tau_{\rm unif}}\;
\end{equation}
implies
\begin{equation}
    \epsilon_\mr{xc}^{\mr{JSCAN}} := \epsilon_\mr{xc}^1 + \left(\epsilon_\mr{xc}^0 - \epsilon_\mr{xc}^1\right)f_\mr{xc}(\widetilde{\alpha}), 
    \label{eq:JSCAN}
\end{equation}
and
\begin{equation}\label{eq:JELF}
{{\rm JELF}} := \frac{1}{1 + \widetilde{\alpha}^2}\;.
\end{equation}
Note that, under the restriction to time-reversal symmetric states, the rest of the SCAN only depends on the particle density (a fully invariant quantity)  and, thus, shall not be modified here. Also note that $\widetilde{\alpha} \ge 0$, as for the original quantity. Importantly, the above substitution should {\em not} be confused with the analogous well-known substitution that involves the (paramagnetic) particle current:\cite{TP05} 
$
    \tau  \to \widetilde{\tau} =\tau -   \frac{\bj \cdot \bj }{2n}.
$
The latter derives from the consideration of local U(1) transformations.  
The importance of the dependence of xc functionals  on the particle current has been already largely demonstrated.
\cite{Dobson93,Becke96,Becke-j02,J-PBE,burnus2005time,TP05,Pittalis07,Pittalis09,Rasanen09,Oliveira2010,Tricky16,Furness2016}

For the states considered in this work as for any gauge-invariant MGGA within GKS,
we stress that
${\vbA}^{_{{\rm JSCAN}}}_\mr{xc}$ does not transform covariantly and does not exert a torque on the spin current.
In fact, from Eq. \eqref{eq:JSCAN}, taking a functional derivative with respect to $\vbJ$, we obtain
\begin{equation}\label{eq:AxcJSCAN}
\frac{1}{c}{\vbA}^{_{{\rm JSCAN}}}_\mr{xc}= -\vbJ \left[ \frac{ \left(\epsilon_\mr{xc}^0 - \epsilon_\mr{xc}^1\right)f'_\mr{xc}(\widetilde{\alpha}) }{ n \tau_{\rm unif}} \right]\;,
\end{equation}
where $f'_\mr{xc}$ denotes the derivative of $f_{xc}$ wrt its own argument.
This is a non-Abelian effective connection; i.e., an xc-spin-vector potential.
This expression shows that ${\vbA}^{_{{\rm JSCAN}}}_\mr{xc}$ is parallel to $\vbJ$ at each point in space and, therefore,
\begin{equation}\label{eq:torque}
\frac{1}{c}{\vA}^{_{{\rm JSCAN}}}_{{\rm xc},\mu} \times \vec{ J}_\mu  = 0\;.
\end{equation}
Yet, the  role of ${\vbA}^{_{{\rm JSCAN}}}_\mr{xc}$ is far from negligible (more below).
The GKS equations for the JSCAN  are reported in the supporting information\cite{ESI_JSCAN} in a form which is handy for numerical implementations.

In summary, Eqs. (\ref{eq:talpha})-(\ref{eq:torque}) are the key equations of this work --- Eqs. \eqref{eq:talpha},  \eqref{eq:AxcJSCAN} and \eqref{eq:torque}  apply to {\em any} MGGA which uses  $\tau$ through  $\alpha$ (for example the TASK functional).\cite{TASK}

\begin{widetext}
\begin{center}
\centering
\begin{figure}[!ht]
\includegraphics[width=\textwidth]{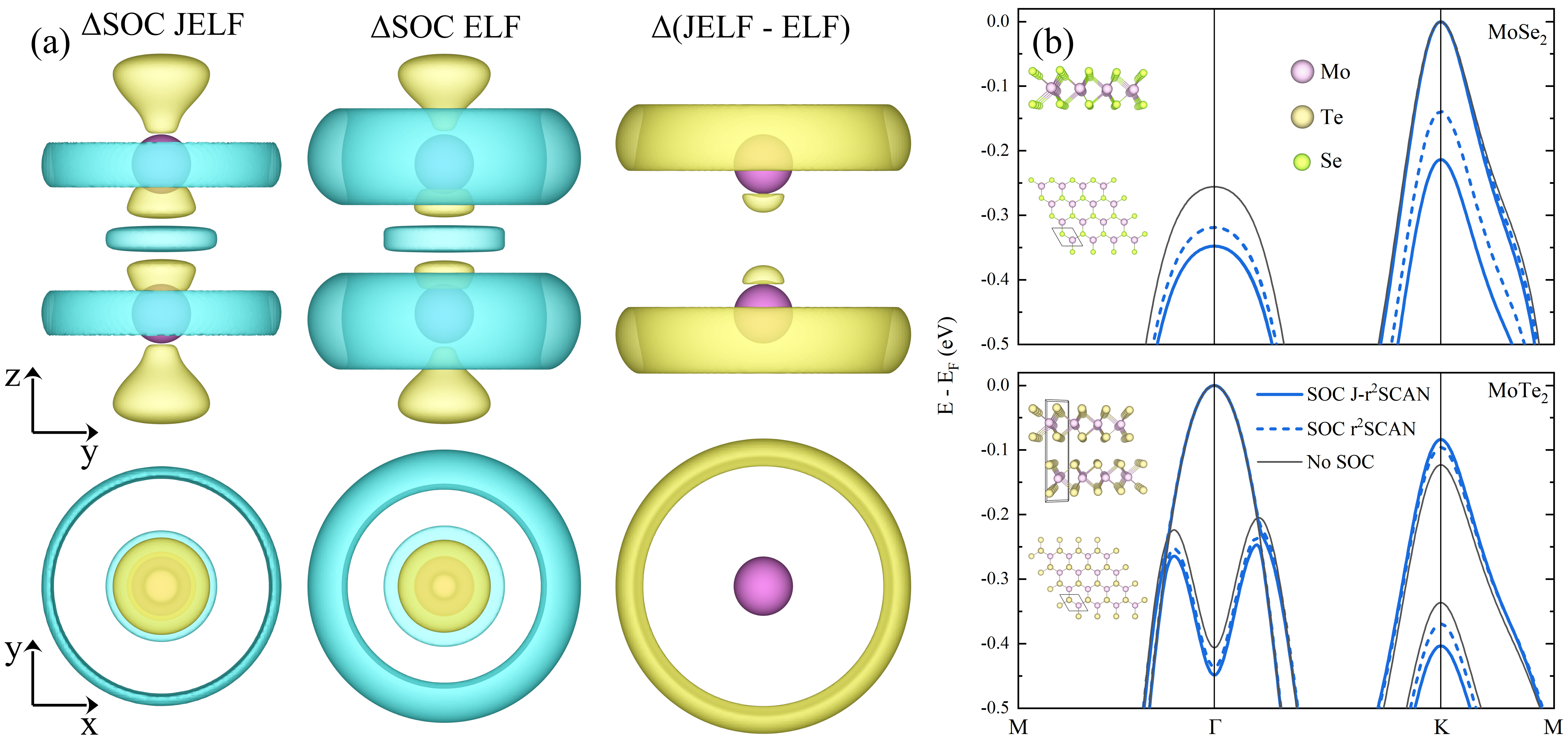}
\caption{(a) Effect of SOC on the ELF and JELF (isosurface at 0.005) in the I$_2$ molecule. $\Delta{\rm SOC}$ (J)ELF is the difference between calculations with and without SOC on the (J)ELF. The rightmost panel reports the difference JELF-ELF (isosurface at 0.002). Positive (negative) isosurfaces are in yellow (blue), while the atoms are represented as purple balls. (b) (top) Valence bands of single-layer MoSe$_2$ and (bottom) bulk $\alpha$-MoTe$_2$ Rashba materials without SOC (black) and with SOC, employing the r$^2$SCAN and J-r$^2$SCAN xc models (dashed and continuous blue lines, respectively).}
\label{fig:I2_JELF}
\end{figure}
\end{center}
\end{widetext}

{\em Applications.}~ 
We have implemented a self-consistent treatment of the JSCAN functional, including effective non-Abelian vector potentials, as well as the JELF in a developer's version of the \textsc{Crystal23} code.\cite{erba23cry} Computational and implementation details are provided in the supporting information\cite{ESI_JSCAN} (see also Refs. \onlinecite{desmarais2021spin2,desmaraistrsb,desmarais2019fundamental,desmarais2019spinI,metz2000small,peterson2003systematically,peterson2007energy,doll2001analytical,doll2001implementation,doll2006analytical,civalleri2001hartree,laun2018consistent,heyd2005energy,lebedev1976quadratures,lebedev1977spherical,towler1996density,Lehtola2020} therein). In the calculations, we employ the restored-regularized r$^2$-SCAN, rather than  the original SCAN functional, to improve numerical stability and performance.\cite{furness2022construction}

\begin{table}[h!]
\caption{We report the effect of SOC on the HOMO-LUMO gap at the level of Exchange-only and Exchange-Correlation using J(SCAN) and the SOC effect on the optical gap 
at the RPAx level.\cite{NWCHEM_NOTE,nwchem2020} The results show that these effects are comparable, and the agreement is improved when spin currents are included in the MGGA. All values are reported in eV.}
\label{tab:gap}
\vspace{2pt}
\begin{tabular}{lccccc}
\hline
\hline
&&\\
&\multicolumn{2}{c}{\textit{Exx-Only}} && \multicolumn{2}{c}{\textit{Exx \& Corr}}\\ 
&&\\
& {\bf J-r$^2$SCAN} & {\bf HF} && {\bf J-r$^2$SCAN} & {\bf RPAx} \\
FI & 0.248 (0.225) & 0.322 && 0.232 (0.222) & 0.360 \\
HI & 0.337 (0.276) & 0.343 && 0.324 (0.281) & 0.343 \\
I$_2$ & 0.333 (0.276) & 0.364 && 0.317 (0.278) & 0.396 \\
HAt & 0.927 (0.675) & 0.891 && 0.884 (0.213) & 0.887 \\
&&\\
\hline
\hline
\end{tabular}
\end{table}

We apply our SU(2) gauge-invariant J-r$^2$SCAN approach to molecules on reproducing the SOC-induced modification of the valence electronic structure in halogen diatoms and hydrides, which have served as model systems in previous studies.\cite{desmarais2021perturbationI,desmarais2021perturbationII,desmarais2023perturbation} 
Values of the SOC-induced modification to the HOMO-LUMO gap are reported in Table \ref{tab:gap}, first at the exchange-only level of theory, and then including effects of exchange and correlation. Values are presented with  exchange-only (gauge-invariant)  J-r$^2$SCAN functional, and numbers in parentheses represent corresponding results with the standard (gauge-dependent) r$^2$SCAN approach.

Inclusion of the spin currents in the functional greatly improves the agreement with Hartree-Fock (HF) results. In the case of the HI molecule, for example, the HOMO-LUMO gap is increased from 0.276 eV to 0.337 eV; which reduces the relative error from 20\% to 2\%. In the case of HAt, the effect of SOC on the HOMO-LUMO gap is increased from 0.675 eV to 0.927 eV; relative error decreased from 24\% to 4\%.

At the xc level of theory, values are benchmarked against the random phase approximation including the dynamical response of Fock exchange (RPAx).\cite{ullrich2011time} In detail, this comparison assumes that the effect of SOC on the fundamental gap is to a good approximation  the same as the effect of SOC on the optical gap. In all cases, inclusion of currents in the xc model improves agreement against the RPAx. The most extreme case is HAt where the effect of SOC on the gap is increased from 0.213 eV to 0.884 eV by inclusion of spin-currents in the MGGA, essentially coinciding with the RPAx value of 0.887 eV.

Next, we discuss  electron localization through the JELF [see Eq. \eqref{eq:JELF}]. We recall that large values of $\tilde{\alpha}$, corresponding to a (J)ELF close to one, indicate that electrons are highly localized. Here we look at the effect of SOC on the (J)ELF. For the I$_2$ molecule, after a self-consistent solution with J-r$^2$SCAN, including both exchange and correlation effects, in Fig. \ref{fig:I2_JELF}(a) we plot the ELF and JELF indicators as differences w.r.t. the calculation without SOC.

Yellow contours indicate that SOC localizes the electron. Blue contours indicate that the electron is delocalized by SOC. The leftmost panel, containing the $\Delta {\rm SOC}$ JELF shows that electrons are localized along the bond axis by SOC. On the other hand, we observe two blue ``onion rings'' around the atomic centers. Thus, SOC localizes electrons along the bond and delocalizes them in the orthogonal directions.
Although this analysis is consistent with previous reports,\cite{desmarais2023structural,gulans2022influence} the degree of delocalization is largely overestimated by the gauge-dependent ELF, compared to the SU(2) form-invariant JELF, see their difference in the rightmost panel of Fig. \ref{fig:I2_JELF}(a).  The JELF displays more localization on the bonding compared to the (gauge-dependent) ELF.

We proceed to apply our approach to the description of the electronic band structure of infinite periodic systems. In Table \ref{tab:gaps}, we provide values of the splitting of the valence band at the $\mathbf{K}$ high-symmetry point in molybdenum dichalcogenide Rashba systems: the inversion asymmetric hexagonal single layer MoSe$_2$ and the inversion-symmetric hexagonal bulk $\alpha$-MoTe$_2$. The valence band structure is provided in panel (b) of Figure \ref{fig:I2_JELF}. 

For both systems, about {\em one third} of the total effect of SOC on the splitting is accounted for through many-body effects by inclusion of spin currents in the functional. Low-temperature experimental values on the transport gap are available from Refs. \onlinecite{choi2017temperature,island2016precise,zelewski2017photoacoustic,conan1979temperature}, while band splittings are provided in Refs. \onlinecite{oliva2020hidden,zhang2014direct,shim2014large,ross2013electrical,beal1972transmission,ruppert2014optical}. The values of valence band splittings are considerably enhanced by including spin-current in the xc form (from 0.14 eV to 0.20 eV in MoSe$_2$, and from 0.28 eV to 0.32 eV in MoTe$_2$), leading to an improved agreement with experiment (relative error decreased from 29\% to 10\% in MoSe$_2$ and leading to an exact coincidence with the mean experimental value in MoTe$_2$). Overall, the accuracy of the J-r$^2$SCAN approach is comparable to more involved hybrid-GGA SCDFT treatments (which provide splittings of 0.17 and 0.32 eV, respectively for MoSe$_2$ and MoTe$_2$ using optimal fractions of Fock exchange) --- {\em without requiring the determination of an optimal fraction of Fock exchange:\cite{desmarais2024generalized} a step that 
involves either empiricism or extra/external  \textit{ab initio} calculations}.\cite{skone2014self}

\begin{table}[h!]
\caption{Splitting of the valence band at $\mathbf{K}$ and band gap of molybdenum dichalcogenide Rashba systems.}
\label{tab:gaps}
\vspace{5pt}
\begin{tabular}{rccccc}
    \hline
    \hline
              & \multicolumn{2}{c}{2D MoSe$_2$} && \multicolumn{2}{c}{3D $\alpha$-MoTe$_2$} \\
    &&\\
        & split & gap && split & gap \\ 
        r$^2$SCAN   & 0.14 & 1.57 && 0.28 & 0.82 \\
        J-r$^2$SCAN & 0.20 & 1.53 && 0.32 & 0.81 \\
        Exp.   &  0.18 & 1.6-2.3 && 0.3-0.34 & 1.03  \\
    \hline
    \hline
    \end{tabular}
\end{table}

Finally, we report on the application of our SU(2) form-invariant approach to topological materials. We consider the case of Weyl semi-metals. We report on the splitting of Weyl nodes in the orthorhombic TaAs phase (W1 node pair, along the $k_z=0$ mirror plane). The pair of nodes is located at $k_y=0.5066$ (in units of  $\pi/\mathbf{b}$), in excellent agreement with the experimental value of $k_y=0.5173$.\cite{lv2015experimental,huang2015weyl} Spin-current dependent terms in the xc functional account for $21\%$ of the total splitting along $k_x$. See Fig. S1 of the supplementary material.

{\em Conclusions}.~ 
Focusing on  the prominent example of the SCAN functional and the ELF, we have shown how to include a dependence on spin currents in meta-GGA functional forms while fulfilling SU(2) gauge invariance for time-reversal symmetric systems.
For the states considered in this work, the current-dependent form of the approximation is furthermore strongly suggested by previous studies on the structure of the exact exchange hole (see Ref. \onlinecite{Pittalis2017}),  which incidentally, also motivate the inclusion of the kinetic energy density as a variable on which the functional must depend.
A large body of works have highlighted the usefulness of the kinetic energy in combination with other quantities as in the ELF (and, thus, as in the iso-orbital indicator $\alpha$ of SCAN)
as a practical means for capturing  relevant local features of the many-electron state in correlated regimes beyond exchange-only effects. 

Of course, the question of including the  ``right physics" beyond the minimalistic yet {\em necessary} approach presented here remains open.
Looking ahead, it will help disentangle the role of the SU(2) invariance in density functional approximations from questions more directly related to the physics of the electron-electron interaction.
It is particularly appealing to explore the consequence of a full (compound) U(1)$\times$SU(2) gauge-invariance on general states that may {\em also} break time-reversal symmetry. This may help solve standing difficulties of present DFT methodologies for magnetized systems.\cite{Isaacs2018,Jana2018,Ekholm2018,Fu2018,Fu2019,Trickey2019,Romero2018,Buchelnikov2019,Tran2020}

\begin{acknowledgments}
This research has received funding from the Project CH4.0 under the MUR program ``Dipartimenti di Eccellenza 2023-2027'' (CUP: D13C22003520001). GV was supported by the Ministry of Education, Singapore, under its Research Centre of Excellence award to the Institute for Functional Intelligent Materials (I-FIM, project No. EDUNC-33-18-279-V12).
\end{acknowledgments}

\appendix


\begin{thebibliography}{107}
\bibitem[{not({\natexlab{b}})}]{note3}
\bibinfo{note}{$\vbG$ is pure gauge vector potential, fully originated by the
  SU(2) transformation}.
\bibitem[{ESI()}]{ESI_JSCAN}
\bibinfo{note}{Computational and implementation details are available at INSERT
  URL HERE}.
\end{thebibliography}

\end{document}